\documentclass[12pt]{article}
\usepackage{color,amssymb,epsfig,verbatim,amsmath}
\def\log{\hbox{log}}

\def\boxit#1{\vbox{\hrule\hbox{\vrule\kern6pt
          \vbox{\kern6pt#1\kern6pt}\kern6pt\vrule}\hrule}}
\def\refhg{\hangindent=20pt\hangafter=1}
\def\refmark{\par\vskip 2mm\noindent\refhg}

\def\refhg{\hangindent=20pt\hangafter=1}
\def\refmark{\par\vskip 2mm\noindent\refhg}

\def\bse{\begin{eqnarray*}}
\def\ese{\end{eqnarray*}}
\def\be{\begin{eqnarray}}
\def\ee{\end{eqnarray}}
\def\bq{\begin{equation}}
\def\eq{\end{equation}}
\def\bse{\begin{eqnarray*}}
\def\ese{\end{eqnarray*}}

\newtheorem{theorem}{Theorem}[section]

\usepackage{lscape}
\usepackage{pdflscape}
\usepackage{afterpage}
\usepackage{longtable}

\usepackage{hyperref}

\hypersetup{
    colorlinks=true,
    linkcolor=blue,
    filecolor=magenta,      
    urlcolor=red,
}

\begin{document}
\thispagestyle{empty}

\hfill\today \\ \\

\baselineskip=28pt
\begin{center}
{\LARGE{\bf Iterated Feature Screening based on Distance Correlation for Ultrahigh-Dimensional Censored Data with Covariates Measurement Error}}
\end{center}
\baselineskip=14pt
\vskip 2mm
\begin{center}
Li-Pang Chen\footnote{\baselineskip=10pt Corresponding Author: Department of Statistics and Actuarial Science, University 
of Waterloo, Waterloo, Ontario, Canada N2L 3G1, L358CHEN@uwaterloo.ca}

\end{center}
\bigskip

\vspace{8mm}

\begin{center}
{\Large{\bf Abstract }}
\end{center}
\baselineskip=17pt
{
Feature screening is an important method to reduce the dimension and capture informative variables in ultrahigh-dimensional data analysis. Many methods have been developed for feature screening. These methods, however, are challenged by complex features pertinent to the data collection as well as the nature of the data themselves. Typically, incomplete response caused by right-censoring and covariates measurement error are often accompanying with survival analysis. Even though there are many methods have been proposed for censored data, little work has been available when both incomplete response and measurement error occur simultaneously. In addition, the conventional feature screening methods may fail to detect the truly important covariates which are marginally independent of the response variable due to correlations among covariates. In this paper, we explore this important problem and propose the valid feature screening method in the presence of survival data with measurement error. In addition, we also develop the iteration method to improve the accuracy of selecting all important covariates. Numerical studies are reported to assess the performance of the proposed method. Finally, we implement the proposed method to two different real datasets.
}

\vspace{8mm}

\par\vfill\noindent

\par\medskip\noindent
\underline{\bf Short title}: Iterated feature screening for censored data and measurement error

\clearpage\pagebreak\newpage
\pagenumbering{arabic}

\newlength{\gnat}
\setlength{\gnat}{22pt}
\baselineskip=\gnat

\clearpage

\section{Introduction} \label{Introduction}

Ultrahigh-dimensional data appears in various scientific research areas, including genetic data, financial data, survival data, and so on. In regression analysis, ultrahigh-dimensional data is very difficult to analyze  since it contains many unimportant variables in the sense that those variables are not highly correlated to the response. In addition, the covariance matrix of the variables is usually singular due to that the dimension of variables is ultra higher than the sample size. As a result, we should select the informative variables before constructing regression models. Moreover, to cope with ultrahigh dimensionality, the assumption of sparsity is imposed. In other words, there are only a small number of predicting variables associated with the response. 

In the early development of variable selection, Akaike's Information Criterion (AIC) (Akaike 1973) and Bayesian Information Criterion (BIC) (Schwarz 1978) are two well-known conventional variable selection criteria. Those two methods aim to search over all possible combinations so that the optimal solution is achieved. However, in ultrahigh-dimensional data, it is near impossible to search the final model through all possible combinations of variables. In the two decades, some regularization methods have been proposed to select variables. Those methods include the LASSO (Tibshirani 1996), SCAD (Fan and Li 2001), LARS (Efron et al. 2004), elastic net (Zou and Hastie 2005), adaptive LASSO (Zou 2006), and Dantzig selector (Candes and Tao 2007) methods. However, those methods are mainly implemented in high-dimensional data but the dimension of variables is smaller than the sample size, and they may perform worse for ultrahigh-dimensional data.

To address ultrahigh-dimensional data with stable computation and accurate selection, Fan and Lv (2008) first proposed the sure independent screening (SIS) procedure for ultrahigh-dimensional linear model which utilized the Pearson correlation to rank the importance of each predictor. Hall and Miller (2009) developed the bootstrap procedure to rank the importance of each predictor based on Pearson correlation between the response and predictors. Fan et al. (2009) and Fan and Song (2010) considered to rank the importance of each predictor through marginal maximum likelihood. Different from the SIS method which specifies the model structure,  Zhu et al. (2011) and Li et al.  (2012) proposed the model-free feature screening to capture the informative covariates for the ultrahigh-dimensional data.

Even though feature screening methods for ultrahigh-dimensional data have been developed, the research gaps still exist. Specifically, in survival analysis with genetic data, the response (failure time) is usually incomplete due to {\it right-censoring} and the covariates are usually contaminated with {\it measurement error}. It is not trivial to implement the conventional feature screening methods to analyze such data. Actually, in the presence of the incomplete response (or survival data) and precise measurement, some valid methods have been proposed. To name a few, Fan et al. (2010) proposed SIS method but restricted on Cox model. Song et al. (2014) proposed the censored rank independence screening. Yan et al. (2017) proposed the Spearman rank correlation screening. Chen et al. (2018) developed the robust feature screening based on distance correlation. Chen et al. (2019) considered a model-free survival conditional feature screening. In the presence of measurement error, however, it is unknown that whether or not those existing methods can determine the ``correct'' features for the surrogate version of the covariates.

The other crucial issue is the {\it accuracy} of feature screening. Since conventional SIS methods rank the importance of each predictor through marginal utilities, then those methods may fail to detect truly important predictors which are marginally independent of the response due to correlations among predictors. The detailed example is deferred in Section~\ref{remark}. To overcome this problem, Fan and Lv (2008) proposed the iterative SIS method. Zhong and Zhu (2015) developed the iterated distance correlation to improve the accuracy of variable screening. These methods, however, are based on complete data and free of mismeasurement. For ultrahigh-dimensional survival data with measurement error in covariates, there is no method to deal with this problem. As a result, we mainly explore this important problem with both survival data and covariates measurement error incorporated. In our development, we first present the distance correlation with error correction for feature screening. Under such approach, the set of selected surrogate variables is the same as the set of selected unobserved variables. After that, we propose the valid iterated procedure with error correction to improve the accuracy of feature screening. In particular, our proposed method is free of model specification and free of specification of distribution for the covariates.

The remainder is organized as follows. In Section~\ref{Notation}, we introduce the survival data with right-censoring, measurement error model and the distance correlation method. In Section~\ref{Method}, we propose the iteration algorithm of feature screening procedure for censored data and covariates measurement error. Empirical studies, including simulation results and real data analysis, are provided in Sections~\ref{Num-study} and \ref{RDA}, respectively. We conclude the article with discussions in Section~\ref{Summary}.

\section{Notation and Model} \label{Notation}

\subsection{Survival Data}  \label{Survival-Data}

In survival analysis, the response $T$ is usually incomplete due to the presence of the censoring time. Specifically, let $T$ be the failure time and $C$ be the censoring time. Then let $Y = \min\{T, C\}$ and denote $\delta = \mathbb{I}(T \leq C)$, where $\mathbb{I}(\cdot)$ is the indicator function. Let $X$ be the $p$-dimensional random vector of covariates. Suppose that we have a sample of $n$ subjects and that for $i=1,\cdots, n$, $(Y_i, \delta_i, X_i)$ has the same distribution as $(Y,\delta,X)$ and $(y_i, \delta_i, x_i)$ represents realizations of $(Y_i, \delta_i, X_i)$. Let $\tau$ denote a maximum support of the failure time. Some regular conditions are imposed.
\begin{itemize}
   \item[(C1)] $P \left( R_i(\tau) = 1 \right)>0$, where $\tau$ is an upper bound of failure times which is assumed to be finite and $R_i(t) = \mathbb{I}\left( Y_i \geq t \right)$ is the risk set.
\item[(C2)] Censoring time is non-informative. That is, the failure time and the censoring time are independent.
\end{itemize}

\subsection{Measurement Error Model}

Let $X^\ast$ denote the surrogate, or observed covariate, of $X$. Let $\Sigma_{X^\ast}$ and $\Sigma_X$ be the covariance matrices of $X^\ast$ and $X$, respectively. For $i=1,\cdots,n$, $(X_i^\ast,X_i)$ has the same distribution as $(X^\ast,X)$. Let $(x_i^\ast,x_i)$ denote the realizations of $(X_i^\ast,X_i)$. In this paper, we focus on the the following measurement error model 
\begin{eqnarray} \label{measure-error-model}
X_i^\ast =  X_i + \epsilon_i
\end{eqnarray}
for $i=1,\cdots,n$, where $\epsilon_i$ is independent to $\left\{X_i, T_i, C_i \right\}$, $\epsilon_i \sim N(0,\Sigma_\epsilon)$ with covariance matrix $\Sigma_\epsilon$. Here $\Sigma_\epsilon$ can be known or unknown. Hence, to discuss $\Sigma_\epsilon$ and its estimation, we consider the following three scenarios:
\begin{description}
\item[Scenario I]: $\Sigma_{\epsilon}$ is known.

In this scenario, $\Sigma_{\epsilon}$ is a constant matrix. Therefore, it is straightforward to discuss the analysis.
\item[Scenario II]: $\Sigma_{\epsilon}$ is unknown and repeated measurements is available. 

Measurement error model (\ref{measure-error-model}) with repeated measurement is given by
\begin{eqnarray*} 
X_{ir}^\ast = X_{i} + \epsilon_{ir}
\end{eqnarray*}
for $i=1,\cdots,n$ and $r=1,\cdots,n_i$, where the $\epsilon_{ir} \sim N \left(0, \Sigma_{\epsilon}\right)$ and independent to $\left\{ \left(X_i, T_i, C_i\right) \right\}$. Using the method of moments, we estimate $\Sigma_\epsilon$ by
\begin{eqnarray} \label{repeat_est_2}
\widehat{\Sigma}_\epsilon = \frac{\sum \limits_{i=1}^n \sum \limits_{r=1}^{n_i} \left(X_{ir}^\ast - \bar{X}_{i \cdot}^\ast \right) \left(X_{ir}^\ast - \bar{X}_{i \cdot}^\ast \right)^\top}{\sum \limits_{i=1}^n \left(n_i - 1 \right)},
\end{eqnarray}
where $\bar{X}_{i\cdot}^\ast = \frac{1}{n_i} \sum \limits_{r=1}^{n_i} X_{ir}^\ast$.
\item[Scenario III]: $\Sigma_{\epsilon}$ is unknown and validation data is available.

Suppose that $\mathcal{M}$ is the subject sets for the main study containing $n$ subjects and $\mathcal{V}$ is the subject sets for the external validation study containing $m$ subjects. Assume that $\mathcal{M}$ and $\mathcal{V}$ do not overlap. Therefore, the
available data contain measurements $\left\{ \left( t_i, c_i, \delta_i, x_i^\ast \right): i \in \mathcal{M} \right\}$ from the main study and $\left\{ \left( x_i, x_i^\ast \right): i \in \mathcal{V} \right\}$ from the validation sample. Hence, for the measurement error model,
we have 
\begin{eqnarray*}
X_i^\ast =  X_i + \epsilon_i
\end{eqnarray*}
for $i \in \mathcal{M} \cup \mathcal{V}$, where the $\epsilon_i \sim N \left(0, \Sigma_\epsilon\right)$ and independent to $\left\{ \left(X_i, T_i, C_i\right) \right\}$. In this case, applying the least square regression method gives
\begin{eqnarray} \label{est: Sigma_epsilon}
\widehat{\Sigma}_\epsilon = \frac{1}{m-1} \sum \limits_{i \in \mathcal{V}} e_i^\top e_i,
\end{eqnarray}
where $e_i = X_i^\ast - X_i$.
\end{description}

\subsection{Review of the Distance Correlation Method} \label{SDR-DC}

In this section, we briefly review the distance correlation (DC) method, which was first proposed by Sz\'ekely et al. (2007).

Let $\phi_\mu(\cdot)$ and $\phi_\nu(\cdot)$ denote the characteristic functions of two random vectors $\mu$ and $\nu$, respectively, and let $\phi_{\mu,\nu}(\cdot)$ be the joint characteristic function of $\mu$ and $\nu$. Let $\left\| \phi(\cdot) \right\|^2 = \phi(\cdot)\bar{\phi}(\cdot)$ for any complex function $\phi(\cdot)$, where $\bar{\phi}(\cdot)$ is the conjugate of $\phi(\cdot)$. The \textit{distance covariance} between $\mu$ and $\nu$ is defined as
\begin{eqnarray*}
\text{dcov}^2(\mu,\nu) = \int_{\mathbb{R}^{d_\mu+ d_\nu}} \left\| \phi_{\mu,\nu}(r,s) -  \phi_\mu(r)\phi_\nu(s) \right\|^2 w(r,s) drds, 
\end{eqnarray*}
where $d_\mu$ and $d_\nu$ are dimensions of $\mu$ and $\nu$, respectively, and 
\begin{eqnarray*}
w(r,s) = \left\{ c_{d_\mu} c_{d_\nu} \left\| r \right\|_{d_\mu}^{1+d_\mu} \left\| s \right\|_{d_\nu}^{1+d_\nu} \right\}^{-1}
\end{eqnarray*}
with $c_d = \pi^{(1+d)/2} / \Gamma \{(1+d)/2\}$ and $\left\| a \right\|_d$ is the Euclidean norm of any vector $a \in \mathbb{R}^d$. Therefore, the DC is defined as
\begin{eqnarray} \label{formula-DC}
\text{dcorr}(\mu,\nu) = \frac{\text{dcov}(\mu,\nu)}{\sqrt{\text{dcov}(\mu,\mu)\text{dcov}(\nu,\nu)}}.
\end{eqnarray}

Sz\'ekely et al. (2007) showed that two random vectors $\mu$ and $\nu$ are independent if and only if $\text{dcorr}(\mu,\nu) = 0$. This property motivates us to do the feature screening and identify which covariates are dependent with the response (e.g., Li et al. 2012). The detailed estimation of (\ref{formula-DC}) can be found in Li et al. (2012).

\subsection{Potential Problem in Conventional Screening Method} \label{remark}

As discussed in Section~\ref{Introduction}, even though many feature screening methods have been proposed, those methods cannot capture all important variables due to that some variables are highly correlated with others. To see this problem explicitly, we consider the following regression model which was adopted by Fan and Lv (2008) 
\begin{equation} \label{Model-example}
T = X_1 + X_2 + X_3 - 3 \sqrt{\rho} X_4 + \epsilon,
\end{equation}
where $X = \left( X_1,\cdots,X_p\right)^\top$ with $p=1000$ is a vector of covariates and each $X_k$ is generated from the normal distribution with mean zero and unit variance. The correlations of all $X_k$ except $X_4$ are $\rho$, while $X_4$ has the correlation $\sqrt{\rho}$ with all other $p-1$ variables.

It is clear to see that variables $X_k$ with $k=1,2,3,4$ are included in model (\ref{Model-example}). By the feature screening based on conventional distance correlation method, we can only identify $X_1, X_2$ and $X_3$, while there is a large probability that $X_4$ cannot be identified due to $\text{dcorr}(X_4,Y) = 0$. 

This simple example verifies that the conventional feature screening method fails to select whole important variables. To successfully identify the variable $X_4$, Fan and Lv (2008) proposed the iterated SIS method. Zhong and Zhu (2015) considered the iterated distance correlation method. In survival analysis, however, the response $T$ in model (\ref{Model-example}) is usually incomplete due to right-censoring. Therefore, in the presence of right-censoring, it is not trivial to implement those conventional methods to deal with survival data. In addition, the other challenge comes from the mismeasurement of covariates. Specifically, variables $X_1$-$X_4$ in model (\ref{Model-example}) may be contaminated with measurement error, and we only have the surrogate variables $X_1^\ast$-$X_4^\ast$. It is expected that the important variables with label 1-4 cannot be identified if we ignore the impact of mismeasurement. As a result, it is also crucial to take care the measurement error effect.

\section{The Proposed Method} \label{Method}

\subsection{Feature Screening for Censored Data and Measurement Error} \label{method-FS}

To present the setting, we start from the unobserved covariate $X$.

Let $F(t|x)$ denote the conditional distribution function of $T$ given $X = \left( X_1,\cdots,X_p\right)^\top$, and let
\begin{eqnarray*}
\mathcal{I} = \left\{ k : F(t|x) \ \text{is functionally dependent to} \ X_k\ \text{for} \ t \in [0,\tau] \right\}
\end{eqnarray*}
denote the \textit{active set} which contains all relevant predictors for the response $T$ with $q = \left|\mathcal{I}\right|$ and $q < n$, and $\mathcal{I}^c$ is the complement of $\mathcal{I}$ which contains all irrelevant predictors for the response $T$. In this case, let $X_{\mathcal{I}} = \left\{X_k : k \in \mathcal{I} \right\}$ denote the vector containing all the active predictors, and let $X_{\mathcal{I}^c} = \left\{X_k : k \in \mathcal{I}^c \right\}$ be the vector containing all the irrelevant predictors.

If $Y$ is complete, i.e., $Y=T$, then it is straightforward to implement conventional methods to determine the active set. However, if $Y$ is incomplete, i.e., right-censoring occurs, then we impute $Y$ by (Buckley and James 1979) 
\begin{eqnarray*}
Y^\ast = \delta Y + (1-\delta) E\left(T | \delta = 0\right),
\end{eqnarray*}
indicating that $E(Y^\ast) = E(T)$ (Miller 1981, p. 151). In addition, by Condition (C1) in Section~\ref{Survival-Data}, $E(T|\delta=0)$ can be written as
\begin{eqnarray*}
E\left(T | \delta = 0\right) 
&=& E\left(T | \tau > T > Y \right) \\
&=& \int_y^\tau t \frac{f_T(t)}{P(\tau > T >y)} dt \\
&=& \int_y^\tau  \frac{tf_T(t)}{1-F_T(y)} dt \\
&=&   \frac{1}{1-F_T(y)} \left[ \left\{ \tau - yF_T(y) \right\}  - \int_y^\tau  F_T(t)   dt \right],
\end{eqnarray*}
where $f_T(\cdot)$ and $F_T(\cdot)$ are the density and distribution functions of $T$, respectively. Moreover, $F_T(\cdot)$ can be estimated by 
\begin{eqnarray*}
\widehat{F}_T(y) = \frac{1}{n} \sum \limits_{i=1}^n \frac{\delta_i}{\widehat{G}(Y_i)} I\left(Y_i \leq y \right), 
\end{eqnarray*}
where $\widehat{G}(y)$ is the Kaplan-Meier estimator of $G(y) = P\left( C\geq y \right)$. As a result, the estimator of $E\left(T | \delta = 0\right)$, denoted as $\widetilde{E\left(T | \delta = 0\right)}$, is determined by replacing $F_T(y)$ with $\widehat{F}_T(y)$, and thus, we have 
\begin{eqnarray*}
\widetilde{Y}^\ast = \delta Y + (1-\delta) \widetilde{E\left(T | \delta = 0\right)}.
\end{eqnarray*}

Finally, the crucial target is the determination of the active set $\mathcal{I}$. In the presence of measurement error, we adopt the DC method described in Section~\ref{SDR-DC} with modification. Let $\phi_Y(r) = E \left\{ \exp \left(\mathbf{i}rY^\ast \right)\right\}$ denote the characteristic function of $Y^\ast$, where $\mathbf{i}$ is a complex number with $\mathbf{i}^2 = -1$. Define
\begin{eqnarray*}
\phi_{X^\ast}(s) = E\left\{ \exp\left(\mathbf{i}s^\top X^\ast \right) \right\} \exp \left( \frac{1}{2} s^\top \Sigma_\epsilon s \right)
\end{eqnarray*}
and
\begin{eqnarray*}
\phi_{Y,X^\ast}(r,s) = E\left\{ \exp\left(\mathbf{i}rY^\ast + \mathbf{i}s^\top X^\ast\right) \right\} \exp \left( \frac{1}{2} s^\top \Sigma_\epsilon s \right).
\end{eqnarray*}
If $\Sigma_\epsilon$ is unknown, then it can be estimated based on repeated measurement or validation in (\ref{repeat_est_2}) or (\ref{est: Sigma_epsilon}). Therefore, we define
\begin{eqnarray*}
\text{dcov}^\ast(Y^\ast,X^\ast) = \int_{\mathbb{R}^{1+p}} \left\| \phi_{Y,X^\ast}(r,s) -  \phi_{Y}(r)\phi_{X^\ast}(s) \right\|^2 w(r,s) drds
\end{eqnarray*}
and
\begin{eqnarray} \label{formula-ModifyDC}
\text{dcorr}^\ast(Y^\ast,X^\ast) = \frac{\text{dcov}^\ast(Y^\ast,X^\ast)}{\sqrt{\text{dcov}^\ast(Y^\ast,Y^\ast)\text{dcov}^\ast(X^\ast,X^\ast)}}.
\end{eqnarray}

As a result, to select features, it suffices to consider
\begin{eqnarray} \label{proposed-feature-selection}
\omega_k = \text{dcorr}^\ast \left(Y^\ast, X_k^\ast \right)
\end{eqnarray}
for $k=1,\cdots,p$, and the corresponding estimator is
\begin{eqnarray}
\widehat{\omega}_k = \widehat{\text{dcorr}^\ast} \left(\widetilde{Y}^\ast, X_k^\ast \right).
\end{eqnarray}
As suggested by Li et al. (2012), let the threshold value be $cn^{-\zeta}$ for some constants $c$ and $\zeta$, then the estimated active set is given by
\begin{eqnarray} \label{estimate-active-set}
\widehat{\mathcal{I}} = \left\{k: \widehat{\omega}_k \geq cn^{-\zeta}, k=1,\cdots,p \right\}.
\end{eqnarray}

To see the validity of the criterion (\ref{proposed-feature-selection}), we have the following theorem:
\begin{theorem} \label{thm-active-set}
Active features based on $X^\ast$ and $X$ are the same. That is, for every $k=1,\cdots,p$,
\begin{eqnarray*}
\text{dcorr}^\ast \left(Y^\ast, X_k^\ast \right) > 0 \Longleftrightarrow \text{dcorr} \left(Y^\ast, X_k \right) > 0
\end{eqnarray*}
or
\begin{eqnarray*}
\text{dcorr}^\ast \left(Y^\ast, X_k^\ast \right) = 0 \Longleftrightarrow \text{dcorr} \left(Y^\ast, X_k \right) = 0,
\end{eqnarray*}
where $\text{dcorr} \left(Y^\ast, X_k \right)$ is determined by implementing $Y^\ast$ and $X_k$ to (\ref{formula-DC}).
\end{theorem} 

Generally speaking, Theorem~\ref{thm-active-set} suggests that based on the feature selection criterion (\ref{proposed-feature-selection}), the true and surrogate covariates share the same active set $\mathcal{I}$. Furthermore, similar derivation in Li et al. (2012) yields that $\widehat{\mathcal{I}}$ has the sure screening property in this sense that $P\left( \mathcal{I} \subseteq \widehat{\mathcal{I}} \right) \rightarrow 1$ as $n \rightarrow \infty$. Therefore, we can decompose the measurement error model (\ref{measure-error-model}) by
\begin{subequations}
\begin{eqnarray}
X_{i,\mathcal{I}}^\ast &=& X_{i,\mathcal{I}} + \epsilon_{i,\mathcal{I}} \label{Decompose-I} \\
X_{i,\mathcal{I}^c}^\ast &=& X_{i,\mathcal{I}^c} + \epsilon_{i,\mathcal{I}^c}, \label{Decompose-Ic}
\end{eqnarray}
\end{subequations}
where $X_i^\ast = \left( X_{i,\mathcal{I}}^{\ast\top}, X_{i,\mathcal{I}^c}^{\ast\top} \right)^\top$, $X_i = \left( X_{i,\mathcal{I}}^{\top}, X_{i,\mathcal{I}^c}^{\top} \right)^\top$, and $\epsilon_i = \left( \epsilon_{i,\mathcal{I}}^{\top}, \epsilon_{i,\mathcal{I}^c}^{\top} \right)^\top$. The covariance matrix $\Sigma_\epsilon$ can be further decomposed as
\begin{eqnarray*}
\Sigma_\epsilon = \left(
\begin{array}{cc}
\Sigma_{\epsilon_\mathcal{I}} & \Sigma_{\epsilon_{\mathcal{I}\mathcal{I}^c}} \\
\Sigma_{\epsilon_{\mathcal{I}\mathcal{I}^c}}^\top & \Sigma_{\epsilon_{\mathcal{I}^c}}
\end{array}
\right),
\end{eqnarray*}
where $\Sigma_{\epsilon_\mathcal{I}}$ is the $q \times q$ covariance matrix based on (\ref{Decompose-I}), $\Sigma_{\epsilon_{\mathcal{I}^c}}$ is the $(p-q) \times (p-q)$ covariance matrix based on (\ref{Decompose-Ic}), and $\Sigma_{\epsilon_{\mathcal{I}\mathcal{I}^c}}$ is the $q \times (p-q)$ covariance matrix based on the  interaction of (\ref{Decompose-I}) and (\ref{Decompose-Ic}).

\subsection{Iteration Algorithm} \label{method-IFS}

As motivated by example in Section~\ref{remark}, directly implementing (\ref{proposed-feature-selection}) may lose some important variables. To increase the probability of selecting all important variables, we modify the selection criterion (\ref{proposed-feature-selection}) and develop the iterated feature screening procedure.

The key idea is as follows: We first implement the feature screening criterion (\ref{proposed-feature-selection}) to determine $X_\mathcal{I}$ and $X_{\mathcal{I}^c}$. It is noted that there exist some potential important variables in $\mathcal{I}^c$ but not be identified. Therefore, to determine the other important variables in  $\mathcal{I}^c$, a natural way is to remove the correlations of $X_{\mathcal{I}^c}$ and $X_{\mathcal{I}}$ by regressing $X_{\mathcal{I}^c}$ onto $X_{\mathcal{I}}$. As a result, the residuals obtained from such linear regression are then uncorrelated with $X_{\mathcal{I}}$. Therefore, other important variables in $\mathcal{I}^c$ can be identified by residuals and $Y^\ast$.

Specifically, to present the idea explicitly, we provide the following iteration algorithm:

\begin{description}

\item[Step 1:] {\it Initial determination of the active set}.

Let $\mathbf{X}^\ast = \left( \mathbf{X}_1^\ast,\cdots,\mathbf{X}_p^\ast \right)$ denote the $(n \times p)$ covariate matrix, where $\mathbf{X}_k^\ast$ is a $n$-dimensional  vector of $k$th covariate with $k=1,\cdots,p$.

In this stage, we first implement (\ref{proposed-feature-selection}) to determine the initial active set $\mathcal{I}$ and the corresponding relevant covariate is $\mathbf{X}_{\mathcal{I}}^\ast$ with dimension $n \times \left| \mathcal{I} \right|$. Let $\mathbf{X}_{\mathcal{I}^c}^\ast$ denote the irrelevant covariates matrix with dimension $n \times \left(p- \left| \mathcal{I} \right| \right)$ such that $\mathbf{X}^\ast = \left( \mathbf{X}_{\mathcal{I}}^\ast, \mathbf{X}_{\mathcal{I}^c}^\ast \right)$. In addition, based on feature selection criterion (\ref{proposed-feature-selection}) and Theorem~\ref{thm-active-set}, the active set based on the surrogate variables is equal to the set based on the true covariates. Therefore, we also have $\mathbf{X} = \left( \mathbf{X}_{\mathcal{I}}, \mathbf{X}_{\mathcal{I}^c} \right)$.

\item[Step 2:] {\it Improvement}.

In this stage, we aim to search other important variables in $\mathcal{I}^c$. Our main approach is to regress $\mathbf{X}_{\mathcal{I}^c}$ onto $\mathbf{X}_{\mathcal{I}}$ and update the active set through the residual.

In this paper, we consider the multivariate linear regression model and the ordinary least square is given by
\begin{eqnarray*}
Q(\beta) \triangleq \left\| \mathbf{X}_{\mathcal{I}^c} - \mathbf{X}_{\mathcal{I}} \beta \right\|^2,
\end{eqnarray*}
where $\left\| \cdot \right\|$ is the $L_2$-norm and $\beta$ is the parameter matrix with dimension $\left| \mathcal{I} \right| \times \left(p - \left| \mathcal{I} \right| \right)$. The corresponding score function is
\begin{eqnarray*}
U(\beta) \triangleq \frac{\partial Q(\beta)}{\partial \beta} \propto  \mathbf{X}_{\mathcal{I}}^\top \left( \mathbf{X}_{\mathcal{I}^c} - \mathbf{X}_{\mathcal{I}} \beta \right).
\end{eqnarray*}
However, in the presence of covariates measurement error , we only observe $\mathbf{X}_{\mathcal{I}}^\ast$ and $\mathbf{X}_{\mathcal{I}^c}^\ast$, then the score function becomes
\begin{eqnarray} \label{naive-score-function}
U^\ast(\beta) \propto  \mathbf{X}_{\mathcal{I}}^{\ast\top} \left( \mathbf{X}_{\mathcal{I}^c}^\ast - \mathbf{X}_{\mathcal{I}}^\ast \beta \right).
\end{eqnarray}

It is well known that directly solving $U^\ast(\beta) = 0$ may incur the estimator of $\beta$ with tremendous bias (e.g., Carroll et al. 2006). Instead, by the simple calculation, we obtain
\begin{eqnarray} \label{correct-score-function}
U^{\ast\ast}(\beta) =  \mathbf{X}_{\mathcal{I}}^{\ast\top}  \mathbf{X}_{\mathcal{I}^c}^\ast - \mathbf{X}_{\mathcal{I}}^{\ast\top} \mathbf{X}_{\mathcal{I}}^\ast \beta - \Sigma_{\epsilon_{\mathcal{I}\mathcal{I}^c}} + \Sigma_{\epsilon_\mathcal{I}} \beta
\end{eqnarray}
such that
\begin{eqnarray*}
E\left\{ U^{\ast\ast}(\beta)  | \mathbf{X} \right\} = U(\beta),
\end{eqnarray*}
indicating that $U^{\ast\ast}(\beta)$ is the suitable score function which corrects error-prone variables. Therefore, the estimator of $\beta$ based on (\ref{correct-score-function}) is given by
\begin{eqnarray} \label{estimator-beta}
\widehat{\beta} = \left( \mathbf{X}_{\mathcal{I}}^{\ast\top} \mathbf{X}_{\mathcal{I}}^\ast - \Sigma_{\epsilon_\mathcal{I}} \right)^{-1} \left( \mathbf{X}_{\mathcal{I}}^{\ast\top}  \mathbf{X}_{\mathcal{I}^c}^\ast - \Sigma_{\epsilon_{\mathcal{I}\mathcal{I}^c}} \right).
\end{eqnarray}

Based on (\ref{estimator-beta}) and the surrogate variables, define
\begin{eqnarray}
\mathbf{X}_\text{new}^\ast = \mathbf{X}_{\mathcal{I}^c}^\ast - \mathbf{X}_{\mathcal{I}}^\ast \widehat{\beta}.
\end{eqnarray}
In fact, $\mathbf{X}_\text{new}^\ast$ is an exact formulation of the residual and thus $\mathbf{X}_\text{new}^\ast$ contains the covariate information in $\mathcal{I}^c$ and is uncorrelated to $\mathbf{X}_{\mathcal{I}}$. Therefore, implementing (\ref{proposed-feature-selection}) with $\mathbf{X}_\text{new}^\ast$ gives the active set $\mathcal{I}_1$ based on $\mathbf{X}_\text{new}^\ast$.

\item[Step 3:] {\it Update of the active set}.

Update the active set $\mathcal{I}$ by $\mathcal{I} \cup \mathcal{I}_1$ and continue Step 2 until no more covariate is included. Finally, the final model is $\mathcal{I}$.

\end{description}

In practice, as suggested in Yan et al. (2017), Chen et al. (2019) and among others, we can specify the size of the active set $\mathcal{I}$ to be $q = \left[ \frac{n}{\log(n)} \right]$, where $\left[\cdot\right]$ stands for the floor function. In this sense, based on the iteration algorithm, we can first select variables with size $q_1 < q$ in Step 1, and then determine the variables with size $q - q_1$ in Step 2.

\section{Simulation Studies} \label{Num-study}

\subsection{Simulation Setup}

Let $n = 400$ denote the sample size. Let $X = \left( X_1,\cdots,X_p\right)^\top$ with $p=2000,4000$, or $6000$ denote a $p$-dimensional vector of covariates which is generated from the normal distribution with mean zero and the covariance matrix $\Sigma_X$ with the diagonal elements being one and the non-diagonal elements being the correlations of all $X_k$ with $k=1,\cdots,p$. Similar to the setting with an example in Section~\ref{remark}, we specify the correlations of all $X_k$ except $X_4$ to be $\rho$, while $X_4$ has the correlation $\sqrt{\rho}$ with all other $p-1$ variables. We consider $\rho = 0.5$ or $0.8$.

The failure time is generated by the following  model:
\begin{eqnarray*}
T = \exp \left( X_1 + X_2 + X_3 - 3 \sqrt{\rho} X_4 + \eta \right).
\end{eqnarray*}
Specifying the distribution of the error term $\eta$ gives some commonly used survival models. In this paper, we consider the extreme value distribution for the proportional hazards (PH) model and the logistic distribution for the proportional odds (PO) model. The censoring time $C$ is generated from the uniform distribution $U(0,\tau_C)$ where $\tau_C$ is a constant such that the censoring rate is approximately 50\%. As a result, we have $Y = \min\{T,C\}$ and $\delta = \mathbb{I}\left(T\leq C \right)$. For $i=1,\cdots,n$, the survival data is $\left( Y_i, \delta_i, X_i \right)$.

For the measurement error model (\ref{measure-error-model}), let $\epsilon_i$ be generated from the normal distribution with mean zero and the diagonal matrix $\Sigma_\epsilon$ with entries being $\sigma_\epsilon^2 = 0.15$, 0.5, or 0.75. If $\Sigma_\epsilon$ is unknown, then the following two scenarios are considered as additional information: 
\begin{description}
\item[Scenario 1:] Repeated measurement

For $i=1,\cdots,n$ and $r=1,\cdots,n_i$ with $n_i=2$, $X_i$ and $\epsilon_{ir}$ are again be generated from $N(0,\Sigma_X)$ and $N(0,\Sigma_\epsilon)$, respectively, and $X_{ir}^\ast$ is generated by
\begin{eqnarray*}
X_{ir}^\ast = X_i + \epsilon_{ir}
\end{eqnarray*}
for $i=1,\cdots,n$ and $r=1,\cdots,n_i$. As a result, $\widehat{\Sigma}_\epsilon$ can be estimated by (\ref{repeat_est_2}).

\item[Scenario 2:] Validation data

For $i=1,\cdots,m$ with $m=100$, $X_i$ and $\epsilon_i$ are again be generated from $N(0,\Sigma_X)$ and $N(0,\Sigma_\epsilon)$, respectively, and $X_i^\ast$ is generated by
\begin{eqnarray*}
X_i^\ast = X_i + \epsilon_i
\end{eqnarray*}
for $i=1,\cdots,m$. Therefore, $\widehat{\Sigma}_\epsilon$ can be estimated by (\ref{est: Sigma_epsilon}).

\end{description}

Finally, we repeat simulation 1000 times in each setting.

\subsection{Simulation Results}

To evaluate the finite-sample performance of the proposed method, we consider the proportion that {\it each active covariates is selected}  out of 1000 simulations which is denoted by $\mathcal{P}_s$, and  the proportion that {\it all active covariates are selected} out of 1000 simulations which is denoted by $\mathcal{P}_a$. In addition, for the comparisons, we also examine the {\it naive estimator}, which is derived by directly implementing the observed covariates  $X_i^\ast$ and taking iteration through (\ref{naive-score-function}). For two different survival models and several settings of $\Sigma_\epsilon$, we compare the results obtained from applying the proposed method to the surrogate covariates as opposed to the estimators obtained from fitting the data with the true covariate measurements.

The numerical results are placed in Tables~\ref{tab:Sim}-\ref{tab:Sim-valid}. Since feature screenings based on the naive and proposed methods use the same criterion (\ref{proposed-feature-selection}), so the screening result are the same. Furthermore,  the results of feature screening based on the true covariates $X$ are similar to the results based on the surrogate covariates $X^\ast$ regardless values of $\rho$ and $\sigma_\epsilon^2$. It also verifies Theorem~\ref{thm-active-set}. However, the feature screening method can successfully select variables $X_1, X_2$ and $X_3$ with high probability, but $X_4$ is selected with low proportion. This result is consistent with the example in Section~\ref{remark}. On the contrary, from Tables~\ref{tab:Sim}-\ref{tab:Sim-valid}, we can see that the iterated feature screening method based on corrected score function (\ref{correct-score-function}) successfully identify the variable $X_4$ with high proportion. This result is parallel to the case that the true covariate is implemented. On the other hand, even the iterated feature screening method is implemented, $X_4$ cannot be identified if the measurement error effect is not corrected appropriately. This result is verified by the naive method with the usage of (\ref{naive-score-function}).

\section{Data Analysis} \label{RDA}

\subsection{Analysis of The Mantle Cell Lymphoma Microarray Data} \label{RDA-DLBCL}

We first illustrate the proposed methods by an application to the mantle cell lymphoma microarray dataset, available from \url{http://llmpp.nih.gov/MCL/}. The dataset contains the survival time of 92 patients and the gene expression measurements
of 8810 genes for each patient. However, we only concern 6312 genes after
deleting 2498 ones appearing to be missing. During the follow-up, 64 patients died of
mantle cell lymphoma and the other 28 ones were censored, causing 36\% censoring ratio. The aim of the study was to formulate a molecular predictor of survival after chemotherapy for the disease.

Since this dataset contains no information to characterize the degree of measurement error that is accompanying with the gene expressions, here we conduct sensitivity analyses to investigate the measurement error effects on analysis results. Specifically, let $\Sigma$ be the covariance matrix of the gene expressions. For sensitivity analyses, we consider $\Sigma + \Sigma_e$ to be the covariance matrix for the measurement error model (\ref{measure-error-model}), where $\Sigma_e$ is the diagonal matrix with diagonal elements being a common value $\sigma_e^2$, which is specified as $\sigma_e^2 = 0.15, 0.55$, or $0.75$ to feature a setting with minor, moderate or severe measurement error. Let $q = \left[ \frac{92}{\log(92)} \right] = 20$, indicating that we aim to select 20 variables in the active set $\mathcal{I}$. In the iteration algorithm, we first select 8 gene expressions, and then the remaining 12 gene expressions are selected by either (\ref{naive-score-function}) or (\ref{correct-score-function}). For comparisons, we examine the feature screening (FS) method in Section~\ref{method-FS} and the iterated feature screening (IFS) method in Section~\ref{method-IFS}. The selection results are summarized in Table~\ref{tab:RDA-DLBCL}.

From Table~\ref{tab:RDA-DLBCL}, we can see that both feature screening and iterated feature screening methods have the same results in the first 8 gene expressions regardless of proposed and naive methods. It indicates that the first 8 gene expressions are clearly dependent on the response and easily identified. In the remaining 12 gene expressions, on the other hand, the screening results are different. Specifically, the iterated feature screening method select some gene expressions, such as 29897, 30620, 32699 and so on, regardless of different degrees of measurement error effects, and those selected gene expressions are not shown in the result of feature screening method. It implies that the iterated feature screening method select some potentially important variables which are not identified by the feature screening method. Furthermore, for the result based on naive method, even the iterated feature screening method is implemented, the selections in the remaining 12 gene expressions are different from the result based on the correction of error effect. The main reason comes from the usage of the estimator of $\beta$ solved by (\ref{naive-score-function}) or (\ref{correct-score-function}).

\subsection{Analysis of NKI Breast Cancer Data}

In this section, we implement our proposed method to analyze the breast cancer data collected by the Netherlands Cancer Institute (NKI) (van de Vijver et al. 2002). Tumors from 295 women with breast cancer were collected from the fresh-frozen-tissue bank of the Netherlands Cancer Institute. Tumors of those patients were primarily invasive breast cancer carcinoma that were about 5 cm in diameter. Patients at diagnosis were 52 years or younger and the  diagnosis was done from 1984 to 1995. Of all those patients, 79 patients died before the study ended, yielding approximately the 73.2\% censoring rate. For each tumor of patient, approximate 25000 gene expressions were collected. Consistent with the analysis of gene expression data, we treat log intensity as the covariates.

Since this dataset also contains no information to characterize the degree of measurement error that is accompanying with the gene expressions, similar to the idea in Section~\ref{RDA-DLBCL}, we conduct sensitivity analyses to investigate the measurement error effects on analysis results. That is, let $\Sigma$ be the covariance matrix of the gene expressions, and we consider $\Sigma + \Sigma_e$ to be the covariance matrix for the measurement error model (\ref{measure-error-model}), where $\Sigma_e$ is the diagonal matrix with diagonal elements being a common value $\sigma_e^2$, which is specified as $\sigma_e^2 = 0.15, 0.55$, or $0.75$ to feature a setting with minor, moderate or severe measurement error. Let $q = \left[ \frac{79}{\log(79)} \right] = 18$, indicating that we aim to select 18 variables in the active set $\mathcal{I}$. In the iteration algorithm, here we first select 7 gene expressions, and then the remaining 11 gene expressions are selected by either (\ref{naive-score-function}) or (\ref{correct-score-function}). Similar to the procedure in Section~\ref{RDA-DLBCL}, we investigate the feature screening (FS) method in Section~\ref{method-FS} and the iterated feature screening (IFS) method in Section~\ref{method-IFS}. The selection results are summarized in Table~\ref{tab:RDA-NKI}.

 From Table~\ref{tab:RDA-NKI}, the result of NKI data is parallel to the result in Section~\ref{RDA-DLBCL}, in the sense that both feature screening and iterated feature screening methods have the same results in the first 7 gene expressions regardless of proposed and naive methods. It indicates that the first 7 gene expressions are clearly dependent on the response and easily identified. In the remaining 11 gene expressions, on the other hand, the screening results are different. For example, the iterated feature screening method select some gene expressions, such as NM\underline{\ \ }020188, Contig25991, NM\underline{\ \ }003882 and so on, regardless of different degrees of measurement error effects, and those selected gene expressions are not shown in the result of feature screening method. It implies that the iterated feature screening method select some potentially important variables which are not identified by the feature screening method. Furthermore, for the result based on naive method, even the iterated feature screening method is implemented, the selections in the remaining 11 gene expressions are different from the result based on the correction of error effect. The main reason comes from the usage of the estimator of $\beta$ solved by (\ref{naive-score-function}) or (\ref{correct-score-function}).

\section{Conclusion} \label{Summary}

Ultrahigh-dimensional data analysis is one of an important topics in decades, and it appears frequently in many practical situations and research fields, such as biological data and financial data. Many methods have been developed to deal with this problem. In the presence of censored data and covariates measurement error simultaneously, however, little method is available. Furthermore, some truly important covariates may be failed to be detected due to correlations among other covariates.

To overcome those challenges, we propose the valid feature screening method to deal with ultrahigh-dimensionality with both censored data and covariates measurement error incorporated simultaneously. Different from other feature screening methods based on censored data, the proposed method enables to determine the same active predictors based on the surrogate and unobserved covariates. To improve the accuracy of feature screening and identify some potentially important variables, we further develop the iterated feature screening with correction of measurement error. Throughout the simulation studies and real data analysis, it is verified that the iterated feature screening method yields the satisfactory results and outperforms the feature screening and naive methods.

There are some possible extensions and applications. First, even the dimension of variables is reduced to be smaller than the sample size, sometimes the dimension is still high and some unimportant variables may still contain in the dataset. In this case, we then implement the variable selection techniques, such as LASSO or SCAD, to identify the most important variables and shrink other unimportant variables. Second, although we mainly consider continuous covariates and classical measurement error model, the proposed method can be naturally extended to other types of variables, such as binary and count variables, and other measurement error models, including Berkson error model. Furthermore, the binary covariates with mismeasurement, also called {\it misclassification}, is also a crucial problem. Finally, in addition to right-censoring, some complex structures, such as left-truncation (e.g., Chen 2019), also appear in survival data with ultrahigh dimensionality. It is also interesting to explore this problem by extending the proposed method. These important topics are our future work.

\appendix
\setcounter{equation}{0}
\renewcommand\theequation{A.\arabic{equation}}
\section*{Appendix: Proof of Theorem~\ref{thm-active-set}}

We first consider $\text{dcov} \left(Y^\ast, X \right)$ and $\text{dcov}^\ast \left(Y^\ast, X^\ast \right)$. Note that the former formulation is based on the true covariates $X$, while the latter formulation is based on the surrogate covariates $X^\ast$.

Since the error term $\epsilon$ follows normal distribution $N(0,\Sigma_\epsilon)$, then its characteristic function is given by
\begin{eqnarray} \label{char-fun-epsilon}
E\left\{ \exp\left( \mathbf{i} s^\top \epsilon \right) \right\} = \exp \left( -\frac{1}{2} s^\top \Sigma_\epsilon s \right).
\end{eqnarray}
By the direct computation, we have
\begin{eqnarray} \label{char-fun-X-ast}
\phi_{X^\ast}(s) &=& E\left\{ \exp\left(\mathbf{i}s^\top X^\ast \right) \right\} \exp \left( \frac{1}{2} s^\top \Sigma_\epsilon s \right) \nonumber \\
&=& E\left\{ \exp\left(\mathbf{i}s^\top X \right)  \right\} E\left\{ \exp\left(\mathbf{i}s^\top \epsilon \right)  \right\} \exp \left( \frac{1}{2} s^\top \Sigma_\epsilon s \right) \nonumber \\
&=& E\left\{ \exp\left(\mathbf{i}s^\top X \right)  \right\}, 
\end{eqnarray}
where the second equality is due to the independence of $X$ and $\epsilon$, and the last equality is due to (\ref{char-fun-epsilon}).

In addition, we can also derive
\begin{eqnarray} \label{char-fun-Y-X-ast}
\phi_{Y,X^\ast}(r,s) &=& E\left\{ \exp\left(\mathbf{i}rY^\ast + \mathbf{i}s^\top X^\ast \right) \right\} \exp \left( \frac{1}{2} s^\top \Sigma_\epsilon s \right) \nonumber \\
&=& E\left\{ \exp\left(\mathbf{i}rY^\ast + \mathbf{i}s^\top X \right) \right\} E\left\{ \exp\left( \mathbf{i}s^\top \epsilon \right) \right\} \exp \left( \frac{1}{2} s^\top \Sigma_\epsilon s \right) \nonumber \\
&=& E\left\{ \exp\left(\mathbf{i}rY^\ast + \mathbf{i}s^\top X \right) \right\},
\end{eqnarray}
where the second equality is due to the independence of $\epsilon$ and $(X, Y)$, and the last equality again comes from (\ref{char-fun-epsilon}). As a result, combining (\ref{char-fun-X-ast}) and (\ref{char-fun-Y-X-ast}) with $\text{dcov}^\ast \left(Y^\ast, X_k^\ast \right)$ gives the same expression of $\text{dcov} \left(Y^\ast, X_k \right)$. 

The equivalence of $\text{dcov}^\ast \left(X_k^\ast, X_k^\ast \right)$ and $\text{dcov} \left(X_k, X_k \right)$ holds by the similar derivations. Therefore, we conclude that $\text{dcorr} \left(Y^\ast, X_k \right)$ and $\text{dcorr}^\ast \left(Y^\ast, X_k^\ast \right)$ are equivalent in the sense that $\text{dcorr} \left(Y^\ast, X_k \right) > 0$ if and only if $\text{dcorr}^\ast \left(Y^\ast, X_k^\ast \right) > 0$. Consequently, the same active features can be determined for $X^\ast$ and $X$. $\hfill \square$

\section*{References}

\refmark Akaike, H. (1973) Information theory and an extension of the maximum likelihood principle. In 2nd {\em International Symposium on Information Theory}, eds by Petrov, N. and Czaki, F., 267 - 281. Akademiai Kaido, Bydapest.

\refmark Buckley, J. and James, I. (1979) Linear regression with censored data. {\em Biometrika}, {66}, 429-436.

\refmark Candes, E. and Tao, T. (2007) The Dantzig selector: statistical estimation when p is much larger than n (with discussion). {\em The Annals of Statistics}, 35, 2313 - 2404.

\refmark Carroll, R. J., Ruppert, D., Stefanski, L. A., and Crainiceanu, C. M. (2006) {\em Measurement Error in Nonlinear Model}. CRC Press, New York.

\refmark Chen, L.-P. (2019). Pseudo likelihood estimation for the additive hazards model with data subject to left-truncation and right-censoring. {\em Statistics and Its Interface}, {\bf 12}, 135-148.

\refmark Chen, X., Chen, X. and Wang, H. (2018) Robust feature screening for ultra-high dimensional right censored data via distance correlation. {\em Computational Statistics and Data Analysis}, { 119}, 118-138.

\refmark Chen, X., Zhang, Y., Chen, X. and Liu, Y. (2019) A simple model-free survival conditional feature screening. {\em Statistics and Probability Letters}, { 146}, 156-160.

\refmark Efron, B., Hastie, T., Johnstone, I. and Tibshirani, R. (2004) Least angle regression. {\em The Annals of Statistics}, 32, 409 - 499.

\refmark Fan, J. and Li, R. (2001) Variable selection via nonconcave penalized likelihood and its oracle properties. {\em Journal of the American Statistical Association}, 96, 1348-1360.

\refmark Fan, J. and Lv, J. (2008) Sure independence screening for ultrahigh dimensional feature space (with discussion). {\em Journal of the Royal Statistical Society. Series B}, 70, 849 - 911.

\refmark Fan, J., Samworth, R. and Wu, Y. (2009) Ultrahigh dimensional feature selection: beyond the linear model. {\em Journal of Machine Learning Research},, 10, 1829 - 1853.

\refmark Fan, J. and Song, R. (2010) Sure independence screening in generalized linear models with NP-dimensionality. {\em The Annals of Statistics}, 38, 3567 - 3604.

\refmark Fan, J., Feng, Y. and Wu, Y. (2010) Ultrahigh dimensional variable selection for Cox's proportional hazards model. IMS Collect, 6, 70 - 86.

\refmark Hall, P. and Miller, H. (2009) Using generalized correlation to effect variable selection in very high dimensional problems.
{\em Journal of Computational and Graphical Statistics}, 18, 533 - 550.

\refmark Li, R., Zhong, W. \& Zhu, L. (2012). Feature screening via distance correlation learning. {\em Journal of the American Statistical Association}, { 107}, 1129 - 1139.

\refmark Miller, R. G. (1981). {\em Survival Analysis}. Wiley, New York.

\refmark Rosenwald, A.,Wright, G., Chan, W. C., Connors, J. M., Campo, E., Fisher, R. I., Gascoyne, R. D., Muller-Hermelink, H. K., Smeland, E. B., and Staudt, L. M. (2002). The use of molecular profiling to predict survival after chemotherapy for diffuse large-B-bell lymphoma. {\em The New England Journal of Medicine}, 346, 1937-1947.

\refmark Schwarz, G. (1978) Estimating the dimension of model. {\em Annals of Statistics}, 6, 461 - 464.

\refmark Sz\'ekely, G. J., Rizzo, M. L. \& Bakirov, N. K. (2007). Measuring and testing dependence by correlation of distances. {\em The Annals of Statistics}, { 35}, 2769-2794.

\refmark Tibshirani, R. (1996) Regression shrinkage and selection via the lasso. {\em Journal of the Royal Statistical Society. Series B}, 58, 267-288.

\refmark van de Vijver, M. J., He, Y. D., van't Veer, L. J., Dai, H., Hart, A. A.M., Voskuil, D. W.,
Schreiber, G.J., Peterse, J.L., Roberts, C., Marton, M.J., Parrish, M., Atsma, D., 
Witteveen, A., Glas, A., Delahaye, L., van der Velde, T., Bartelink, H., 
Rodenhuis, S., Rutgers, E.T., Friend, S.H. and Bernards, R. (2002) A gene-expression signature as a predictor of survival in breast cancer. {\em The New England Journal of Medicine}, 347, 1999 - 2009.

\refmark Yan, X., Tang, N. and Zhao, X. (2017) The Spearman rank correlation screening for ultrahigh dimensional censored data. arXiv:1702.02708v1

\refmark Zhong, W. and Zhu, L. (2015) An iterative approach to distance correlation-based sure independence screening. {\em Journal of Statistical Computation and Simulation}, 85, 2331 - 2345.

\refmark Zhu, L., Li, L., Li, R. and Zhu, L. (2011). Model-free feature screening for ultrahigh-dimensional data. {\em Journal of the American Statistical Association}, { 106}, 1464 - 1475.

\refmark Zou, H. and Hastie, T. (2005) Regularization and variable selection via the elastic net. {\em Journal of Royal Statistical Society: Series B}, 67, 301-320.

\refmark Zou, H. (2006) The adaptive lasso and its oracle properties. {\em Journal of the American Statistical Association}, 101, 1418-1429.

\clearpage

 \begin{table}
       \huge
     \caption{Simulation results for the feature selection with known $L$ and $p=2000$} \label{tab:Sim}

   \scriptsize

 \centering
 
 \begin{tabular}{c c c c c ccccc  c ccccc} 
 \\
 \hline\hline
  &  &   &  & & \multicolumn{5}{c} { Feature screening } & & \multicolumn{5}{c} { Iterated feature screening }\\ \cline{6-10} \cline{12-16} 
 &  &   &  & & \multicolumn{4}{c} { $\mathcal{P}_s$ } & $\mathcal{P}_a$ & & \multicolumn{4}{c} { $\mathcal{P}_s$ } & $\mathcal{P}_a$\\ \cline{6-9} \cline{12-15} 
Model & $\rho$ & $\sigma_{\epsilon}^2$ & Method & &  $X_1$ & $X_2$ & $X_3$ & $X_4$ & & & $X_1$ & $X_2$ & $X_3$ & $X_4$ &
\\
 \hline 
PH &  $0.5$ & 0.15 & Naive &  & 1.000 & 1.000 & 1.000 & 0.005 & 0.005 & & 1.000 & 1.000 & 1.000 & 0.006 & 0.006 \\
  &     &   & Propose &  & 1.000 & 1.000 & 1.000 & 0.005 & 0.005 & & 1.000 & 1.000 & 1.000 & 1.000 & 1.000 \\ \cline{3-16}
  &  & 0.50 & Naive &  & 1.000 & 1.000 & 1.000 & 0.005 & 0.005 & & 1.000 & 1.000 & 1.000 & 0.006 & 0.006
 \\
  &    &   & Propose &  & 1.000 & 1.000 & 1.000 & 0.005 & 0.005 & & 1.000 & 1.000 & 1.000 & 0.998 & 0.998
 \\ \cline{3-16}
  &   & 0.75 & Naive &  & 1.000 & 1.000 & 1.000 & 0.004 & 0.004 & & 1.000 & 1.000 & 1.000 & 0.006 & 0.006
 \\
   &  &   & Propose &  & 1.000 & 1.000 & 1.000 & 0.004 & 0.004 & & 1.000 & 1.000 & 1.000 & 0.996 & 0.996
 \\ \cline{3-16}
  &  & $\times$ & $-$ &  & 1.000 & 1.000 & 1.000 & 0.005 & 0.005 & & 1.000 & 1.000 & 1.000 & 1.000 & 1.000
 \\  
    \\
 & $0.8$ & 0.15 & Naive &  &   1.000 & 1.000 & 1.000 & 0.005 & 0.005 & & 1.000 & 1.000 & 1.000 & 0.006 & 0.006
\\
   &   &   & Propose &  &  1.000 & 1.000 & 1.000 & 0.004 & 0.004 & & 1.000 & 1.000 & 1.000 & 0.996 & 0.996 \\ \cline{3-16}
 &  & 0.50 & Naive &  &  1.000 & 1.000 & 1.000 & 0.005 & 0.005 & & 1.000 & 1.000 & 1.000 & 0.006 & 0.006 \\
   &   &   & Propose &  & 1.000 & 1.000 & 1.000 & 0.004 & 0.004 & & 1.000 & 1.000 & 1.000 & 0.996 & 0.996 \\ \cline{3-16}
   &  & 0.75 & Naive &  &   1.000 & 1.000 & 1.000 & 0.005 & 0.005 & & 1.000 & 1.000 & 1.000 & 0.006 & 0.006
\\
   &    &   & Propose &  & 1.000 & 1.000 & 1.000 & 0.004 & 0.004 & & 1.000 & 1.000 & 1.000 & 0.996 & 0.996 \\ \cline{3-16}
    &  & $\times$ & $-$ &  & 1.000 & 1.000 & 1.000 & 0.005 & 0.005 & & 1.000 & 1.000 & 1.000 & 0.997 & 0.997 \\  
    \\
PO &  $0.5$ & 0.15 & Naive &  & 1.000 & 1.000 & 1.000 & 0.005 & 0.005 & & 1.000 & 1.000 & 1.000 & 0.006 & 0.006  \\
    &   &   & Propose &  & 1.000 & 1.000 & 1.000 & 0.004 & 0.004 & & 1.000 & 1.000 & 1.000 & 0.996 & 0.996 \\ \cline{3-16}
 &   & 0.50 & Naive &  & 1.000 & 1.000 & 1.000 & 0.005 & 0.005 & & 1.000 & 1.000 & 1.000 & 0.006 & 0.006 \\
  &     &   & Propose &  & 1.000 & 1.000 & 1.000 & 0.004 & 0.004 & & 1.000 & 1.000 & 1.000 & 0.996 & 0.996 \\ \cline{3-16}
 &    & 0.75 & Naive &  & 1.000 & 1.000 & 1.000 & 0.005 & 0.005 & & 1.000 & 1.000 & 1.000 & 0.006 & 0.006 \\
  &     &   & Propose &  & 1.000 & 1.000 & 1.000 & 0.004 & 0.004 & & 1.000 & 1.000 & 1.000 & 0.996 & 0.996 \\ \cline{3-16}
   &   & $\times$ & $-$ &  & 1.000 & 1.000 & 1.000 & 0.005 & 0.005 & & 1.000 & 1.000 & 1.000 & 1.000 & 1.000 \\  
       \\
 &  $0.8$ & 0.15 & Naive &  &  1.000 & 1.000 & 1.000 & 0.005 & 0.005 & & 1.000 & 1.000 & 1.000 & 0.006 & 0.006 \\
    &   &   & Propose &  & 1.000 & 1.000 & 1.000 & 0.004 & 0.004 & & 1.000 & 1.000 & 1.000 & 0.996 & 0.996 \\ \cline{3-16}
 &   & 0.50 & Naive &  & 1.000 & 1.000 & 1.000 & 0.005 & 0.005 & & 1.000 & 1.000 & 1.000 & 0.006 & 0.006 \\
  &     &   & Propose &  &  1.000 & 1.000 & 1.000 & 0.004 & 0.004 & & 1.000 & 1.000 & 1.000 & 0.996 & 0.996\\ \cline{3-16}
 &    & 0.75 & Naive &  & 1.000 & 1.000 & 1.000 & 0.005 & 0.005 & & 1.000 & 1.000 & 1.000 & 0.006 & 0.006 \\
  &     &   & Propose &  & 1.000 & 1.000 & 1.000 & 0.004 & 0.004 & & 1.000 & 1.000 & 1.000 & 0.996 & 0.996 \\ \cline{3-16}
   &   & $\times$ & $-$ &  & 1.000 & 1.000 & 1.000 & 0.002 & 0.002 & & 1.000 & 1.000 & 1.000 & 0.997 & 0.997 \\  
 \hline\hline
\end{tabular}
\end{table}

 \begin{table}
       \huge
     \caption{Simulation results for the feature selection with repeated measurement and $p=4000$} \label{tab:Sim-repeat}

   \scriptsize

 \centering
 
 \begin{tabular}{c c c c c ccccc  c ccccc} 
 \\
 \hline\hline
  &  &   &  & & \multicolumn{5}{c} { Feature screening } & & \multicolumn{5}{c} { Iterated feature screening }\\ \cline{6-10} \cline{12-16} 
 &  &   &  & & \multicolumn{4}{c} { $\mathcal{P}_s$ } & $\mathcal{P}_a$ & & \multicolumn{4}{c} { $\mathcal{P}_s$ } & $\mathcal{P}_a$\\ \cline{6-9} \cline{12-15} 
Model & $\rho$ & $\sigma_{\epsilon}^2$ & Method & &  $X_1$ & $X_2$ & $X_3$ & $X_4$ & & & $X_1$ & $X_2$ & $X_3$ & $X_4$ &
\\
 \hline 
PH &  $0.5$ & 0.15 & Naive &  & 1.000 & 1.000 & 1.000 & 0.005 & 0.005 & & 1.000 & 1.000 & 1.000 & 0.004 & 0.004 \\
  &     &   & Propose &  & 1.000 & 1.000 & 1.000 & 0.005 & 0.005 & & 1.000 & 1.000 & 1.000 & 1.000 & 1.000 \\ \cline{3-16}
  &  & 0.50 & Naive &  & 1.000 & 1.000 & 1.000 & 0.005 & 0.005 & & 1.000 & 1.000 & 1.000 & 0.005 & 0.005
 \\
  &    &   & Propose &  & 1.000 & 1.000 & 1.000 & 0.005 & 0.005 & & 1.000 & 1.000 & 1.000 & 0.997 & 0.997
 \\ \cline{3-16}
  &   & 0.75 & Naive &  & 1.000 & 1.000 & 1.000 & 0.004 & 0.004 & & 1.000 & 1.000 & 1.000 & 0.004 & 0.004
 \\
   &  &   & Propose &  & 1.000 & 1.000 & 1.000 & 0.004 & 0.004 & & 1.000 & 1.000 & 1.000 & 0.997 & 0.997
 \\ \cline{3-16}
  &  & $\times$ & $-$ &  & 1.000 & 1.000 & 1.000 & 0.005 & 0.005 & & 1.000 & 1.000 & 1.000 & 1.000 & 1.000
 \\  
    \\
 & $0.8$ & 0.15 & Naive &  &   1.000 & 1.000 & 1.000 & 0.004 & 0.004 & & 1.000 & 1.000 & 1.000 & 0.005 & 0.005
\\
   &   &   & Propose &  &  1.000 & 1.000 & 1.000 & 0.004 & 0.004 & & 1.000 & 1.000 & 1.000 & 0.995 & 0.995 \\ \cline{3-16}
 &  & 0.50 & Naive &  &  1.000 & 1.000 & 1.000 & 0.003 & 0.003 & & 1.000 & 1.000 & 1.000 & 0.002 & 0.002 \\
   &   &   & Propose &  & 1.000 & 1.000 & 1.000 & 0.003 & 0.003 & & 1.000 & 1.000 & 1.000 & 0.994 & 0.994 \\ \cline{3-16}
   &  & 0.75 & Naive &  &   1.000 & 1.000 & 1.000 & 0.004 & 0.004 & & 1.000 & 1.000 & 1.000 & 0.006 & 0.006
\\
   &    &   & Propose &  & 1.000 & 1.000 & 1.000 & 0.004 & 0.004 & & 1.000 & 1.000 & 1.000 & 0.996 & 0.996 \\ \cline{3-16}
    &  & $\times$ & $-$ &  & 1.000 & 1.000 & 1.000 & 0.006 & 0.006 & & 1.000 & 1.000 & 1.000 & 0.998 & 0.998 \\  
    \\
PO &  $0.5$ & 0.15 & Naive &  & 1.000 & 1.000 & 1.000 & 0.003 & 0.003 & & 1.000 & 1.000 & 1.000 & 0.004 & 0.004  \\
    &   &   & Propose &  & 1.000 & 1.000 & 1.000 & 0.003 & 0.003 & & 1.000 & 1.000 & 1.000 & 0.997 & 0.997 \\ \cline{3-16}
 &   & 0.50 & Naive &  & 1.000 & 1.000 & 1.000 & 0.003 & 0.003 & & 1.000 & 1.000 & 1.000 & 0.004 & 0.004 \\
  &     &   & Propose &  & 1.000 & 1.000 & 1.000 & 0.003 & 0.003 & & 1.000 & 1.000 & 1.000 & 0.995 & 0.995 \\ \cline{3-16}
 &    & 0.75 & Naive &  & 1.000 & 1.000 & 1.000 & 0.003 & 0.003 & & 1.000 & 1.000 & 1.000 & 0.004 & 0.004 \\
  &     &   & Propose &  & 1.000 & 1.000 & 1.000 & 0.003 & 0.003 & & 1.000 & 1.000 & 1.000 & 0.995 & 0.995 \\ \cline{3-16}
   &   & $\times$ & $-$ &  & 1.000 & 1.000 & 1.000 & 0.005 & 0.005 & & 1.000 & 1.000 & 1.000 & 1.000 & 1.000 \\  
       \\
 &  $0.8$ & 0.15 & Naive &  &  1.000 & 1.000 & 1.000 & 0.002 & 0.002 & & 1.000 & 1.000 & 1.000 & 0.003 & 0.003 \\
    &   &   & Propose &  & 1.000 & 1.000 & 1.000 & 0.002 & 0.002 & & 1.000 & 1.000 & 1.000 & 0.997 & 0.997 \\ \cline{3-16}
 &   & 0.50 & Naive &  & 1.000 & 1.000 & 1.000 & 0.003 & 0.003 & & 1.000 & 1.000 & 1.000 & 0.006 & 0.006 \\
  &     &   & Propose &  &  1.000 & 1.000 & 1.000 & 0.003 & 0.003 & & 1.000 & 1.000 & 1.000 & 0.995 & 0.995\\ \cline{3-16}
 &    & 0.75 & Naive &  & 1.000 & 1.000 & 1.000 & 0.002 & 0.002 & & 1.000 & 1.000 & 1.000 & 0.003 & 0.003 \\
  &     &   & Propose &  & 1.000 & 1.000 & 1.000 & 0.002 & 0.002 & & 1.000 & 1.000 & 1.000 & 0.995 & 0.995 \\ \cline{3-16}
   &   & $\times$ & $-$ &  & 1.000 & 1.000 & 1.000 & 0.002 & 0.002 & & 1.000 & 1.000 & 1.000 & 0.997 & 0.997 \\  
 \hline\hline
\end{tabular}
\end{table}

 \begin{table}
       \huge
     \caption{Simulation results for the feature selection with validation data and $p=6000$} \label{tab:Sim-valid}

   \scriptsize

 \centering
 
 \begin{tabular}{c c c c c ccccc  c ccccc} 
 \\
 \hline\hline
  &  &   &  & & \multicolumn{5}{c} { Feature screening } & & \multicolumn{5}{c} { Iterated feature screening }\\ \cline{6-10} \cline{12-16} 
 &  &   &  & & \multicolumn{4}{c} { $\mathcal{P}_s$ } & $\mathcal{P}_a$ & & \multicolumn{4}{c} { $\mathcal{P}_s$ } & $\mathcal{P}_a$\\ \cline{6-9} \cline{12-15} 
Model & $\rho$ & $\sigma_{\epsilon}^2$ & Method & &  $X_1$ & $X_2$ & $X_3$ & $X_4$ & & & $X_1$ & $X_2$ & $X_3$ & $X_4$ &
\\
 \hline 
PH &  $0.5$ & 0.15 & Naive &  & 1.000 & 1.000 & 1.000 & 0.007 & 0.007 & & 1.000 & 1.000 & 1.000 & 0.007 & 0.007 \\
  &     &   & Propose &  & 1.000 & 1.000 & 1.000 & 0.007 & 0.007 & & 1.000 & 1.000 & 1.000 & 1.000 & 1.000 \\ \cline{3-16}
  &  & 0.50 & Naive &  & 1.000 & 1.000 & 1.000 & 0.005 & 0.005 & & 1.000 & 1.000 & 1.000 & 0.005 & 0.005
 \\
  &    &   & Propose &  & 1.000 & 1.000 & 1.000 & 0.005 & 0.005 & & 1.000 & 1.000 & 1.000 & 0.997 & 0.997
 \\ \cline{3-16}
  &   & 0.75 & Naive &  & 1.000 & 1.000 & 1.000 & 0.003 & 0.003 & & 1.000 & 1.000 & 1.000 & 0.004 & 0.003
 \\
   &  &   & Propose &  & 1.000 & 1.000 & 1.000 & 0.003 & 0.003 & & 1.000 & 1.000 & 1.000 & 0.995 & 0.995
 \\ \cline{3-16}
  &  & $\times$ & $-$ &  & 1.000 & 1.000 & 1.000 & 0.005 & 0.005 & & 1.000 & 1.000 & 1.000 & 1.000 & 1.000
 \\  
    \\
 & $0.8$ & 0.15 & Naive &  &   1.000 & 1.000 & 1.000 & 0.005 & 0.005 & & 1.000 & 1.000 & 1.000 & 0.005 & 0.005
\\
   &   &   & Propose &  &  1.000 & 1.000 & 1.000 & 0.005 & 0.005 & & 1.000 & 1.000 & 1.000 & 0.997 & 0.997 \\ \cline{3-16}
 &  & 0.50 & Naive &  &  1.000 & 1.000 & 1.000 & 0.004 & 0.004 & & 1.000 & 1.000 & 1.000 & 0.004 & 0.004 \\
   &   &   & Propose &  & 1.000 & 1.000 & 1.000 & 0.004 & 0.004 & & 1.000 & 1.000 & 1.000 & 0.996 & 0.996 \\ \cline{3-16}
   &  & 0.75 & Naive &  &   1.000 & 1.000 & 1.000 & 0.001 & 0.001 & & 1.000 & 1.000 & 1.000 & 0.006 & 0.006
\\
   &    &   & Propose &  & 1.000 & 1.000 & 1.000 & 0.001 & 0.001 & & 1.000 & 1.000 & 1.000 & 0.994 & 0.994 \\ \cline{3-16}
    &  & $\times$ & $-$ &  & 1.000 & 1.000 & 1.000 & 0.005 & 0.005 & & 1.000 & 1.000 & 1.000 & 0.998 & 0.998 \\  
    \\
PO &  $0.5$ & 0.15 & Naive &  & 1.000 & 1.000 & 1.000 & 0.008 & 0.008 & & 1.000 & 1.000 & 1.000 & 0.009 & 0.009  \\
    &   &   & Propose &  & 1.000 & 1.000 & 1.000 & 0.008 & 0.008 & & 1.000 & 1.000 & 1.000 & 0.998 & 0.998 \\ \cline{3-16}
 &   & 0.50 & Naive &  & 1.000 & 1.000 & 1.000 & 0.005 & 0.005 & & 1.000 & 1.000 & 1.000 & 0.005 & 0.005 \\
  &     &   & Propose &  & 1.000 & 1.000 & 1.000 & 0.005 & 0.005 & & 1.000 & 1.000 & 1.000 & 0.997 & 0.997 \\ \cline{3-16}
 &    & 0.75 & Naive &  & 1.000 & 1.000 & 1.000 & 0.005 & 0.005 & & 1.000 & 1.000 & 1.000 & 0.006 & 0.006 \\
  &     &   & Propose &  & 1.000 & 1.000 & 1.000 & 0.005 & 0.005 & & 1.000 & 1.000 & 1.000 & 0.996 & 0.996 \\ \cline{3-16}
   &   & $\times$ & $-$ &  & 1.000 & 1.000 & 1.000 & 0.006 & 0.006 & & 1.000 & 1.000 & 1.000 & 1.000 & 1.000 \\  
       \\
 &  $0.8$ & 0.15 & Naive &  &  1.000 & 1.000 & 1.000 & 0.003 & 0.003 & & 1.000 & 1.000 & 1.000 & 0.004 & 0.004 \\
    &   &   & Propose &  & 1.000 & 1.000 & 1.000 & 0.003 & 0.003 & & 1.000 & 1.000 & 1.000 & 0.995 & 0.995 \\ \cline{3-16}
 &   & 0.50 & Naive &  & 1.000 & 1.000 & 1.000 & 0.002 & 0.002 & & 1.000 & 1.000 & 1.000 & 0.004 & 0.004 \\
  &     &   & Propose &  &  1.000 & 1.000 & 1.000 & 0.002 & 0.002 & & 1.000 & 1.000 & 1.000 & 0.995 & 0.995\\ \cline{3-16}
 &    & 0.75 & Naive &  & 1.000 & 1.000 & 1.000 & 0.000 & 0.000 & & 1.000 & 1.000 & 1.000 & 0.003 & 0.003 \\
  &     &   & Propose &  & 1.000 & 1.000 & 1.000 & 0.000 & 0.000 & & 1.000 & 1.000 & 1.000 & 0.994 & 0.994 \\ \cline{3-16}
   &   & $\times$ & $-$ &  & 1.000 & 1.000 & 1.000 & 0.003 & 0.003 & & 1.000 & 1.000 & 1.000 & 0.998 & 0.998 \\  
 \hline\hline
\end{tabular}
\end{table}

\clearpage

\begin{landscape}
 \begin{table}
       \huge
     \caption{Sensitivity Analyses of mantle cell lymphoma microarray dataset. FS stands for feature screening method in Section~\ref{method-FS}; IFS stands for iterated feature screening method in Section~\ref{method-IFS}.} \label{tab:RDA-DLBCL}

   \scriptsize

 \centering
 
 \begin{tabular}{c c c c c c ccccc  c ccccc} 
 \\
 \hline\hline
 
\# & \multicolumn{2}{c} { $\sigma_e^2 = 0.15$ } & & \multicolumn{2}{c} { $\sigma_e^2 = 0.55$ } & & \multicolumn{2}{c} { $\sigma_e^2 = 0.75$ } & & \multicolumn{2}{c} { naive } \\ \cline{2-3} \cline{5-6} \cline{8-9} \cline{11-12}
& FS & IFS & & FS & IFS & & FS & IFS & & FS & IFS \\ 
 \hline 
1 & 16587 & 16587 & & 16587 & 16587 & & 16587 & 16587 & & 16587 & 16587 \\
2 & 24719 & 24719 & & 24719 & 24719 & & 24719 & 24719 & & 24719 & 24719 \\
3 & 27057 & 27057 & & 27057 & 27057 & & 27057 & 27057 & & 27057 & 27057 \\
4 & 28581 & 28581 & & 28581 & 28581 & & 28581 & 28581 & & 28581 & 28581 \\
5 & 31420 & 31420 & & 31420 & 31420 & & 31420 & 31420 & & 31420 & 31420 \\
6 & 34790 & 34790 & & 34790 & 34790 & & 34790 & 34790 & & 34790 & 34790 \\
7 & 28581 & 28581 & & 28581 & 28581 & & 28581 & 28581 & & 28581 & 28581 \\
8 &  16312 & 29357 & & 16312 & 29357 & & 16312 & 29357 & & 30157 & 30157\\
9 & 34771 & 29897 & & 26537 & 29897 & & 17053
 & 29897 & & 27116 & 28872\\
10 & 28346 & 30620 & & 29637
 & 30620 & & 30917 & 30620 & & 30334 & 32699 \\
11 & 26521
 & 30898 & & 16587 & 30898 & & 30929 & 30898 & & 27762 & 27095\\
12 & 34375 & 32699 & & 17053
 & 32699 & & 31972 & 32699 & & 17326 & 24710\\
13 & 29642
 & 15843 & & 28346 & 15843 & & 29637 & 15844 & & 27019 & 19325\\
14 & 26537 & 15924 & & 28908 & 15924 & & 17605 & 15924 & & 27762 & 30282\\
15 & 17605 & 27927 & & 32519 & 27927 & & 28346 & 27931 & & 17176 & 32187\\
16 & 28920 & 28929 & & 26521 & 28929 & & 34771 & 28929 & & 23887 & 29209\\
17 & 29657 & 34339 & & 34364 & 34375 & & 28908 & 34375 & & 17343 & 16528\\
18 & 32519 & 34913 & & 34667 & 32475 & & 34651 & 32475 & & 32699 & 27019\\
19 & 34651 & 26510 & & 34771 & 26510 & & 16079 & 26510 & & 30157 & 23887\\
20 & 28908 & 27530 & & 27931 & 34913 & & 26537  & 27530 & & 17917 & 16020 \\
 \hline\hline
\end{tabular}
\end{table}
\end{landscape}

\clearpage

\begin{landscape}
 \begin{table}
       \huge
     \caption{Sensitivity Analyses of NKI data. FS stands for feature screening method in Section~\ref{method-FS}; IFS stands for iterated feature screening method in Section~\ref{method-IFS}. } \label{tab:RDA-NKI}

   \scriptsize

 \centering
 
 \begin{tabular}{c c c c c c ccccc  c ccccc} 
 \\
 \hline\hline
 
\# & \multicolumn{2}{c} { $\sigma_e^2 = 0.15$ } & & \multicolumn{2}{c} { $\sigma_e^2 = 0.55$ } & & \multicolumn{2}{c} { $\sigma_e^2 = 0.75$ } & & \multicolumn{2}{c} { naive } \\ \cline{2-3} \cline{5-6} \cline{8-9} \cline{11-12}
& FS & IFS & & FS & IFS & & FS & IFS & & FS & IFS \\ 
 \hline 
1 & NM\underline{\ \ }016359 & NM\underline{\ \ }016359 & & NM\underline{\ \ }016359 &NM\underline{\ \ }016359 & & NM\underline{\ \ }016359 & NM\underline{\ \ }016359 & & NM\underline{\ \ }016359 & NM\underline{\ \ }016359 \\
2 & AA555029\underline{\ \ }RC & AA555029\underline{\ \ }RC & & AA555029\underline{\ \ }RC & AA555029\underline{\ \ }RC & & AA555029\underline{\ \ }RC & AA555029\underline{\ \ }RC & & AA555029\underline{\ \ }RC & AA555029\underline{\ \ }RC \\
3 & NM\underline{\ \ }003748  & NM\underline{\ \ }003748 & & NM\underline{\ \ }003748  & NM\underline{\ \ }003748 & & NM\underline{\ \ }003748  & NM\underline{\ \ }003748 & & NM\underline{\ \ }003748  & NM\underline{\ \ }003748  \\
4 & Contig38288\underline{\ \ }RC  & Contig38288\underline{\ \ }RC  & & Contig38288\underline{\ \ }RC  & Contig38288\underline{\ \ }RC  & & Contig38288\underline{\ \ }RC  & Contig38288\underline{\ \ }RC  & & Contig38288\underline{\ \ }RC  & Contig38288\underline{\ \ }RC  \\
5 & NM\underline{\ \ }003862 & NM\underline{\ \ }003862 & & NM\underline{\ \ }003862 & NM\underline{\ \ }003862 & & NM\underline{\ \ }003862 & NM\underline{\ \ }003862 & & NM\underline{\ \ }003862 & NM\underline{\ \ }003862 \\
6 & Contig28552\underline{\ \ }RC  & Contig28552\underline{\ \ }RC  & & Contig28552\underline{\ \ }RC  & Contig28552\underline{\ \ }RC  & & Contig28552\underline{\ \ }RC  & Contig28552\underline{\ \ }RC  & & Contig28552\underline{\ \ }RC  & Contig28552\underline{\ \ }RC  \\
7 & Contig32125\underline{\ \ }RC & Contig32125\underline{\ \ }RC & & Contig32125\underline{\ \ }RC & Contig32125\underline{\ \ }RC & & Contig32125\underline{\ \ }RC & Contig32125\underline{\ \ }RC & & Contig32125\underline{\ \ }RC & Contig32125\underline{\ \ }RC \\
8 & AB037863 & Contig036649\underline{\ \ }RC & & Contig55725\underline{\ \ }RC & Contig036649\underline{\ \ }RC & & Contig55725\underline{\ \ }RC & Contig036649\underline{\ \ }RC & & NM\underline{\ \ }000599 & NM\underline{\ \ }000599\\
9 & Contig036649\underline{\ \ }RC & Contig46218\underline{\ \ }RC & & AF201905 & Contig46218\underline{\ \ }RC & & AB037863 & Contig46218\underline{\ \ }RC & & Contig46223 & NM\underline{\ \ }005915\\
10 & X05610 & AB037863 & & AB037863 & AB037863 & & AF201905 & AB037863 & & AF257175 & Contig46223 \\
11 & AL080079 & NM\underline{\ \ }020188 & & Contig48328\underline{\ \ }RC & NM\underline{\ \ }020188 & & Contig036649\underline{\ \ }RC & NM\underline{\ \ }020188 & & NM\underline{\ \ }006931 & X05610 \\
12 & NM\underline{\ \ }006931 & Contig55377\underline{\ \ }RC & & Contig036649\underline{\ \ }RC & Contig25991  & & X05610 & Contig25991  & & AK000745 & AK000745\\
13 & AF201905 & Contig48328\underline{\ \ }RC & & AL080079 & Contig55377\underline{\ \ }RC & & NM\underline{\ \ }018354 & Contig48328\underline{\ \ }RC & & NM\underline{\ \ }005915 & NM\underline{\ \ }005915\\
14 & NM\underline{\ \ }003875 & Contig25991  & & X05610 & Contig46223\underline{\ \ }RC & & AL080079 & Contig55377\underline{\ \ }RC & &NM\underline{\ \ }001282 & NM\underline{\ \ }001282\\
15 & Contig55725\underline{\ \ }RC & NM\underline{\ \ }003875 & & Contig55725\underline{\ \ }RC & NM\underline{\ \ }003875 & & Contig55725\underline{\ \ }RC & NM\underline{\ \ }003875 & & AL080079 & NM\underline{\ \ }614321\\
16 & Contig48328\underline{\ \ }RC & NM\underline{\ \ }006101  & & NM\underline{\ \ }018354 & NM\underline{\ \ }006101  & & NM\underline{\ \ }006931 & NM\underline{\ \ }006101 & & NM\underline{\ \ }014889 & AF257175\\
17 & NM\underline{\ \ }000599 & NM\underline{\ \ }003882 & & NM\underline{\ \ }003875 & NM\underline{\ \ }003607 & & Contig48328\underline{\ \ }RC & NM\underline{\ \ }000849  & & Contig55725\underline{\ \ }RC & NM\underline{\ \ }014889\\
18 & NM\underline{\ \ }018354 &  NM\underline{\ \ }016577 & & NM\underline{\ \ }006931 & NM\underline{\ \ }003882 & & NM\underline{\ \ }003875 &  NM\underline{\ \ }016577 & &NM\underline{\ \ }614321 & AF201905 \\
 \hline\hline
\end{tabular}
\end{table}
\end{landscape}

\end{document}